\documentclass[12pt,prd,letter,preprint,nofootinbib]{revtex4-1}
\usepackage{amsmath,amssymb}
\begin{document}

\title{{\bf Gauged WZW models for space-time groups and gravitational actions}}
\author{Pablo Mora$^1$, Pablo Pais$^1$,
and Steven Willison$^2$
\\
$^1${\it Instituto de F\'{\i}sica, Facultad de Ciencias}\\
{\it Igu\'a 4225, Montevideo, Uruguay}\\
$^2${\it Centro de Estudios Cient\'{\i}ficos (CECs)}\\
{\it Casilla 1469, Valdivia, Chile}}

\begin{abstract}
In this paper we investigate gauged Wess-Zumino-Witten models for space-time groups as gravitational theories, following the trend of recent work by Anabalon, Willison and Zanelli. We discuss the field equations in any dimension and study in detail the simplest case of two space-time dimensions and gauge group SO(2,1).
For this model we study black hole solutions and we calculate their mass and entropy which resulted in a null value for both.\\


\end{abstract}
\maketitle

\section{Introduction}

The construction of theories of gravity that are also true gauge theories has lead to the study of Chern-Simons gravities, first in 2+1 dimensions \cite{achucarro,chern-simons-2+1-2}, where these theories are equivalent, at least modulo the question of invertibility of the metric, to General Relativity, and then in higher dimensions \cite{chamseddine-1,chamseddine-2}. Chern-Simons gravities have been extended to the supersymmetric case in refs.\cite{troncoso-zanelli-1,troncoso-zanelli-2}. These theories have a wealth of interesting dynamical properties and solutions including black holes, that have been studied in many papers\footnote{for a recent review and an extensive and comprehensive list of references see ref.\cite{zanelli-lectures}}. Chern-Simons theories are defined in odd-dimensional manifolds and therefore some kind of dimensional reduction or compactification would be required to describe the observed four dimensional universe.

Chern-Simons gauge and gravity theories have been extended by using transgression forms instead of Chern-Simons forms as actions \cite{potsdam,borowiec-1,borowiec-2,IRS-1,IRS-2,sarda-1,sarda-2,tesis,motz1,motz3}, which makes the action gauge invariant instead of just quasi-invariant and has the advantage, in the case of gravity, that the boundary terms required to regularize the conserved charges and black hole thermodynamics are built in \cite{motz1,motz3}.

Wess-Zumino-Witten (WZW) and gauged wess-Zumino-Witten (gWZW) theories \cite{wess-zumino,witten-gWZW} were first introduced as effective theories in nuclear and particle physics. WZW and gWZW models are closely related to Chern-Simons and transgression theories, as they can be regarded as induced at the boundary of a manifold in which CS or transgresion theories respectively are defined if a given gauge field is pure gauge. WZW and gWZW theories are defined in even-dimensional manifolds.

Recently, a particular type of gWZW models for space-time groups have been considered as gravitational theories by Anabalon et al.\cite{anabalon-1,anabalon-2}, whom furthermore shown that in 3+1 dimensions, for the gauge group SO(4,2) and with some additional assumptions the model yields the field equations of General Relativity with cosmological constant. The particular model used is the G-G model (in which the full diagonal subgroup is gauged) which is explicitly topological (without the usual kinetic term).

As Chern-Simons gravities, gWZW models for space-time groups provide gauge theories of gravity, that could be relevant for instance for the construction of a consistent quantum gravity and as modified gravity theories that could solve in a dynamical way problems of modern cosmology such as the nature of dark energy or the cause of inflation in the early universe (if an infationary phase did indeed occur). Some possible problems with this approach are: the emergence of the geometrical interpretation of certain components of the fields which are a priori simply fields living in a topological space; the interpretation of the other field components (maybe as matter or dark energy, etc.).

In this paper we study these gravitational gWZW models along the lines of refs.\cite{anabalon-1,anabalon-2}, focusing in a simple two-dimensional toy model, as a first step towards a future more profound investigation of higher dimensional models of possible phenomenological interest.

In section 2 we review some background material and then derive the field equations of generic (not just gravitational) topological G-G gWZW models in any dimension.

In section 3 we give the explicit field equations of the gWZW model in two dimensions and for gauge group $SU(2)$, as a warm up exercise for the model with gauge group SO(2,1), in which we are interested. In section 4 we derive black hole solutions and discuss the Noether mass and thermodynamics of those solutions. Surprisingly, at least at first glance, the mass and entropy of the black hole turn out to be zero.

\section{Transgressions and gauged WZW actions}

\subsection{Transgressions}

Chern-Simons forms $\mathcal{C}_{2n+1}(A)$ are differential forms defined for
a connection $A$, which under gauge transformations of that connection transform by a closed form, so they are said to be \textit{quasi invariant}. Transgression forms $\mathcal{T}_{2n+1}$ are a generalization of Chern-Simons forms which are strictly invariant. Transgressions depend on two connections, $A$ and $\overline{A}$, and can be written as the difference of two Chern-Simons forms plus an exact form
$$
\mathcal{T}_{2n+1}=\mathcal{Q}_{2n+1}(A)-\mathcal{Q}_{2n+1}(\overline
{A})-dB_{2n}\left(  A,\overline{A}\right),
$$
or also as\footnote{In what follows wedge product between forms is implicitly assumed.} (for the mathematical definitions and properties of Chern-Simons and transgression forms see \cite{nakahara,zumino-les-houches,alvarez} and references therein),
\begin{equation}
\mathcal{T}_{2n+1}\left(  A,\overline{A}\right)  =(n+1)\int_{0}^{1}%
dt\ <\Delta{A}F_{t}^{n}>,
\end{equation}
where
$A_{t} = tA+(1-t)\overline{A}=\overline{A}+t\Delta A $
is a connection that interpolates between the two independent gauge potentials
$A$ and $\overline{A}$ and $\Delta A=A-\overline{A}$. The Lie algebra-valued one-forms  $A=A_{\mu}^{A}G_{A}\ dx^{\mu}$ and $\overline{A}=\overline{A}_{\mu}^{A}G_{A}\ dx^{\mu}$ are
connections under gauge transformations, $G_{A}$ are the generators of the Lie algebra of the gauge group $G$, and
$<\cdots>$ stands for a symmetrized invariant trace in the Lie algebra\footnote{Greek indices are used as space-time indices with values from $0$ to $d-1$ where $d$ is the dimensionality of spacetime; lower case Latin indices from the beginning of the alphabet $a,~b,~c,...$ are tangent space (or Lorentz) indices with values from $0$ to $d-1=2n$; lower case Latin indices from the middle of the alphabet $i,~j,~k,...$ will be used as ordinary vector indices with values from $0$ to $3$. Upper case Latin indices label the generators $G_{A}$ of the Lie group considered and take values from 1 to the dimension of the group.}.
The curvatures (or field strengths) are $F=dA+A^2$, $\overline{F}=d\overline{A}+\overline{A}^2$, while for the interpolating connection the curvature is
$F_{t}=dA_{t}+A_{t}^{2}=t F +(1-t)\overline{F}+t(t-1)(\Delta A)^{2}$.  Setting $\overline{A} =0$
in the transgression form yields the Chern-Simons form $\mathcal{Q}_{2n+1}(A)$ for $A$, that is $\mathcal{Q}_{2n+1}(A)\equiv \mathcal{T}_{2n+1}(A,\overline{A}=0)$.

The exterior derivative of the transgression form gives globally the difference between the invariant polynomials associated to each gauge connection
$$<F^{n+1}>-<\overline{F}^{n+1}>=d\mathcal{T}_{2n+1}(A,\overline{A}).$$

Transgressions have been used to define gauge invariant actions for field theories that generalize Chern-Simons actions in refs.\cite{potsdam,borowiec-1,borowiec-2,IRS-1,IRS-2,sarda-1,sarda-2,tesis,motz1,motz3}. For those actions $A$ and $\overline{A}$ may be taken as defined in distinct manifolds $\mathcal{M}$ and $\overline{\mathcal{M}}$ respectively, with a common boundary $\partial\mathcal{M}\equiv \partial\overline{\mathcal{M}}$, and it is possible either to consider both fields as independent dynamical fields or to consider $\overline{A}$ as a fixed (non dynamical) background. The action for those theories is
$$
I_{Trans}(A,\overline{A})=\int _{\mathcal{M}}\mathcal{Q}_{2n+1}(A)-\int _{\overline{\mathcal{M}}}\mathcal{Q}_{2n+1}(\overline
{A})-\int _{\partial\mathcal{M}}B_{2n}\left(  A,\overline{A}\right).
$$
Those actions generalize Chern-simons actions, providing in the case of Chern-Simons gravity the boundary terms necessary to have a well defined action principle and regularize the conserved charges and black hole thermodynamics \cite{motz1,motz3}.

\subsection{Gauged Wess-Zumino-Witten actions}

A class of related but different theories is obtained if $A$ and $\overline{A}$ are taken to be not independent, but related by a gauge transformation generated by an element $h(x)$ of $G$
$$\overline{A}=h^{-1}Ah+h^{-1}dh\equiv A^h$$
and the manifolds $\mathcal{M}$ and $\overline{\mathcal{M}}$ are the same.

Then the degrees of freedom of the theory correspond to $A$ and $h$. The difference between the Chern-Simons form for $A^h$ and the Chern-Simons form for $A$ is given by the sum of an exact and a closed form \cite{zumino-les-houches}
$$\mathcal{Q}_{2n+1}(A^h)-\mathcal{Q}_{2n+1}(A)=d\alpha _{2n}+\mathcal{Q}_{2n+1}(h^{-1}dh)$$
where $\mathcal{Q}_{2n+1}(h^{-1}dh)$ is the Wess-Zumino-Witten (WZW) form, which satisfies $d\mathcal{Q}_{2n+1}(h^{-1}dh)=0$.

The gauged Wess-Zumino-Witten (gWZW) action is defined as
\begin{equation}
I_{gWZW}(A,h)=\int _{\mathcal{M}}\mathcal{T}_{2n+1}(A,A^h)
\end{equation}
which can be written as
$$
I_{gWZW}(A,h)=\int _{\mathcal{M}}[\mathcal{Q}_{2n+1}(A)-\mathcal{Q}_{2n+1}(
A^h)]-\int _{\partial\mathcal{M}}B_{2n}\left(  A,A^h\right)
$$
or equivalently
$$
I_{gWZW}(A,h)=-\int _{\mathcal{M}}\mathcal{Q}_{2n+1}(h^{-1}dh) -\int _{\partial\mathcal{M}}[\alpha _{2n}+B_{2n}\left(  A,A^h\right)]
$$
The action $I_{gWZW}$ is effectively $2n$-dimensional and lives at the boundary $\partial\mathcal{M}$, for even though the WZW part lives in the $2n+1$-dimensional bulk $\mathcal{M}$, the way in which it is extended into the bulk is immaterial at the quantum level provided that the constant in front of the action (included in the definition of the invariant symmetrized trace) is quantized \cite{witten-gWZW}, and we will see below that the bulk is irrelevant for the classical field equations derived from the action.

The action $I_{gWZW}(A,h)$ is invariant under local gauge transformations generated by a point dependent element $g$ of the group $G$ given by $A\rightarrow g^{-1}[A+d]g$ and $h\rightarrow g^{-1}hg$, as it follows from its definition as a transgression.

For instance the gWZW action in 1+1 dimensions is \cite{anabalon-1}
\begin{equation}\label{action}
I_{gWZW}=\kappa\int_{\Sigma}{\frac{1}{3}\langle(h^{-1}dh)^{3}\rangle}-\kappa\int_{M^{2}}{\langle(A-h^{-1}dh)A^{h}\rangle}
\end{equation}
where $M^2\equiv\partial\Sigma$ is the space-time of the theory.

\subsection{Field equations for the gWZW action in any dimension}

We derived the field equations in any dimension  from the  variation of the action, which was computed starting from the formula for the variation of the transgression \cite{motz3}
$$
\delta\mathcal{T}_{2n+1}=(n+1) <F^n\delta A>-<\overline{F}^n\delta \overline{A}>
-n(n+1)d\{\int _0^1dt<\Delta AF_t^{n-1}\delta A_t>\}
$$
For the detailed derivation see Appendix A.
The  resulting variation of the gWZW action is (though see next subsection for discussion of a subtle point)
\begin{eqnarray}\label{variation_action}
\delta I_{gWZW}=-(n+1)\int _{\Sigma ^{2n}}\Bigg\{
n \int _0^1dt~t<[(A-A^h)F_t^{n-1}+(A^{h^{-1}}-A)\tilde{F}_t^{n-1}]\delta A>+\Bigg.\\ \nonumber
\Bigg. +\int _0^1dt<F_t^{n}h^{-1}\delta h>+n\int _0^1dt~t(t-1)<\Delta A F_t^{n-1}[\Delta A,h^{-1}\delta h]>\Bigg\}+\\ \nonumber
+n(n+1)\int _{\partial\Sigma ^{2n}}\Big\{\int _0^1dt~t<\Delta A F_t^{n-1}h^{-1}\delta h>\Big\}
\end{eqnarray}
where the space-time manifold is $\Sigma ^{2n}\equiv\partial\mathcal{M}$ and $\partial\Sigma ^{2n}$ is its $2n-1$-dimensional boundary, and $A^h =h^{-1}(A+d)h$, $A^{h^{-1}}=h(A+d)h^{-1}$, $\Delta A =A-A^h$, $A_t=tA+(1-t)A^h$, $\tilde{A}_t=t~A+(1-t)A^{h^{-1}}$, $F_t=dA_t+A_t^2$ and $\tilde{F}_t=d\tilde{A}+\tilde{A}_t^2$.

From this expression we can read the field equations derived from the action principle $\delta I_{gWZW}=0$, with the ones corresponding to $\delta A$ being
\begin{eqnarray}
\int _0^1dt~t<[(A-A^h)F_t^{n-1}+(A^{h^{-1}}-A)\tilde{F}_t^{n-1}]G^A>=0
\end{eqnarray}
and the ones corresponding to $\delta h$ being
\begin{eqnarray}
 \int _0^1dt<F_t^{n}G^A>+n\int _0^1dt~t(t-1)<\Delta A F_t^{n-1}[\Delta A,G^A]>=0
\end{eqnarray}
In the particular case of two dimensions ($n=1$), and assuming that the matrix $M^{AB}=<G^AG^B>$ is invertible
it results
\begin{eqnarray}
A^h-A^{h^{-1}}=0\\
F+F^h-\Delta A ^2=F+F^h-\frac{1}{2}[A^h-A,A^h-A]=0
\end{eqnarray}
This equations can be rewritten for later use as
\begin{eqnarray}
D(h^2)=0\label{1stequnrewritten}\\
F+F^h-(h^{-1}Dh)^2=0\label{2ndequnrewritten}
\end{eqnarray}
because $A^h-A^{h^{-1}}=h^{-1}D(h^2)h^{-1}$ and $A^h-A=h^{-1}Dh$.
Finally, note that (\ref{2ndequnrewritten}) can also be rewritten as:
\begin{eqnarray}
 2F + D(h^{-1}Dh) =0\, .
\end{eqnarray}

\subsection{Boundary terms and action principle}

In principle,  $I_{WZW}$ is defined as the integral of the transgression in a manifold with boundary, with that boundary being the space-time in which we are interested. However, in describing physical spacetime, we are interested in the case that $M_2$ is noncompact, for example with topology of $\mathbb{R}^2$. In this case we may use (\ref{action}) as our action, provided we regard $\Sigma$ to be a manifold whose boundary is the topological sphere $M_2 \cup \{\infty\}$.
This makes sense under appropriate asymptotic conditions on $h$ as one approaches infinity, i.e. $h$ is asymptotically constant. The black hole solution of vanishing mass considered below satisfies this condition. We emphasise that he one-point compactification is only for the purpose of defining the Wess-Zumino term for $h$- the gauge field $A$ may have nontrivial behavour at infinity.

The second term in (\ref{action}) may be regarded as an integral over a spacetime with a boundary at infinity. Therefore the variation of the action will produce a boundary contribution, which is precisely the last term in equation (\ref{variation_action}), specialised to $n=1$. This term is proportional to $\delta h$ and so the extremal action principle is consistent with the boundary conditions on $h$.

\section{Field equations in two dimensions}

We shall mainly be interested in the group SO(2,1), with its interpretation as the (anti)-de Sitter spacetime group in 1+1 dimensions. We shall also briefly consider the group SU(2) as a warmup.
But first let us make some general observations, which are valid for and subgroup of GL(2,$\mathbb{C}$) (this is of relevance since the component of SO(2,1) which is connected to the identity is isomorphic to SL(2,$\mathbb{R}$)).

Field equations (\ref{1stequnrewritten}) and (\ref{2ndequnrewritten}) simplify greatly if we assume that the connection is trivial. Setting $A=0$
the equations reduce to $
 d(h^2) =0$, $dh \wedge dh =0$.
Let $h = \left(\begin{smallmatrix} A &
B\\C&D\end{smallmatrix}\right)$ be a complex matrix of nonvanishing
determinant. Then $d(h^2)=0$ implies:
\begin{align*}
 2AdA + B dC + CdB =0\, ,
 \\
 B d(A+D) + (A+D) dB =0 \,
 \\
 C d(A+D) + (A+D) dC =0 \,
 \\
 2DdD + B dC + CdB =0\, .
\end{align*}
First we consider the case $A+D \neq 0$. Then using the above
equations, we can always obtain $\det (h_0) \, dA =0$ or $\det
(h_0)\, dD =0$, which then implies that all the components are
constants. Then we consider the traceless case $A+D =0$. This means
$h_0^2 = -\det(h_0)\, I$. The above equations are solved if
$\det(h_0)$ is constant, with no further restriction on $A,B,C,D$.
So to summarise, we have two types of matrices:
i) $h_0$ is constant;
ii) $h_0$ is a traceless matrix of constant determinant which may have nontrivial degrees of freedom, subject to $dh\wedge dh =0$.
Therefore, there is a special class of matrices, which for group SU(2) or SL(2,$\mathbb{R}$) satisfy $h^2 = -I$ which have especially rich structure.

The above applies in the case of a globally flat connection. A less trivial special case is when $F +F^h =0$. This also simplifies the field equations. In this case, although the calculations are more complicated, it seems that traceless matrices are also special, in that they allow a rich structure of solutions for the field $h$. It is these class of solutions which we shall focus on in what follows.

\subsection{$SU(2)$ Group}\label{su_2}
We begin by deriving the field equations for a simple case: the $SU(2)$ group. In order to derive that equations explicitly we must start with some algebraic preliminaries.

The Pauli matrices are
$
\sigma ^1=\left(\begin{array}{cc}
0 & 1 \\
1 & 0
\end{array}\right)
$,
$
\sigma ^2=\left(\begin{array}{cc}
0 & -i \\
i & 0
\end{array}\right)
$,
$
\sigma ^3=\left(\begin{array}{cc}
1 & 0 \\
0 & -1
\end{array}\right)
$ while the identity is $
I=\left(\begin{array}{cc}
1 & 0 \\
0 & 1
\end{array}\right)
$,
satisfying $\sigma ^i\sigma ^j=\delta ^{ij}I+i\epsilon ^{ijk}\sigma ^k$ and hence
$\{\sigma ^i,\sigma ^j\}=2\delta ^{ij}I$ and $[\sigma ^i,\sigma ^j]=2i\epsilon ^{ijk}\sigma ^k$.
It also follows that $\sigma ^i\sigma ^j\sigma ^k=i\epsilon ^{ijk}I+\delta ^{ij}\sigma ^k+\delta ^{jk}\sigma ^i-\delta ^{ik}\sigma ^j$.

A generic element of SU(2)is of the form
$h=h^0I+h^i\sigma ^i=h^0I+{\bf h.\sigma }$
with the constraint $(h^0)^2+h^ih^i=1$ or equivalently $(h^0)^2+\mid {\bf h}\mid ^2=1$. If  $h$ is written in exponential form it is
$h=e^{i\theta{\bf n.\sigma}}=\cos\theta +i\sin\theta {\bf n.\sigma}$
with ${\bf n.n}=1$.

The gauge potential (or connection) matrix valued one-form is $A=-i A^i\sigma ^i=-i{\bf A.\sigma }$
while the field strength (or curvature) matrix valued two-form is $F=dA+A^2=-iF^i\sigma ^i=-i{\bf F.\sigma }$.
It follows $F^i=dA^i+\epsilon ^{ijk} A^jA^k $ or
$ {\bf F} = d{\bf A}+ {\bf A\times A}$.
If we have an algebra valued zero-form field $\alpha =-i\alpha ^i\sigma ^i=-i{\bf \alpha .\sigma}$ its covariant derivative is $D\alpha =-iD{\bf \alpha}.{\bf \sigma}=d\alpha +[A,\alpha ]$ or $D{\bf \alpha}=d{\bf \alpha}+2{\bf A\times \alpha}$

The equation $A^h-A^{h^{-1}}=0$ yields
\begin{equation}
h^0[d{\bf h}-2 {\bf h\times A}]-{\bf h}dh^0=0
\end{equation}
or equivalently
\begin{equation}
h^0D{\bf h}={\bf h}dh^0
\end{equation}
The equation $F+F^h-\frac{1}{2}[A^h-A,A^h-A]=0$ yields
\begin{equation}
{\bf F}+{\bf h}\times ({\bf h\times F})+h^0({\bf h\times F})-\frac{1}{2}({\bf h}\times D{\bf h})\times ({\bf h}\times D{\bf h})=0
\end{equation}

\subsection{$SO(2,1)$ Group}\label{so_21}

If we want this theory to behave as a simple toy gravity model, we can use the $SO(2,1)$ group. We make use of the conventions in the Section \ref{su_2}.

The generators of SO(2,1) are $J_{ab}$ with $a ,b =0,1,2$. One can relabel those generators as $J^{a}=\epsilon ^{abc}J_{bc}$, and $J_{a}=\eta _{ab}J^{b}$, with $\eta_{ab}=(-,+,+)$. Then the SO(2,1) algebra is written
$$[J_{a},J_{b}]=-2\epsilon _{abc}J^{c}$$
A convenient representation of this algebra is $J^0=-i\sigma_3$, $J^1=-\sigma _1$ and $J^2=-\sigma _2$. For this representation the following useful relations hold
$$J_{a}J_{b}=\eta _{ab}I-\epsilon _{abc}J^{c}$$
and hence
\begin{eqnarray}
\{J_{a},J_{b} \}=2\eta _{ab}I\nonumber\\
J_{a}J_{b}J_{c}=-\epsilon _{abc}+\eta _{ab}J_{c}-\eta _{ac}J_{b}+\eta _{cb}J_{a}
\end{eqnarray}
The gauge potential (connection) is $A=A^{a}J_{a}$, while an element of the group is of the form $h=\lambda I+\alpha ^{a}J_{a}$, with $\lambda ^2-\alpha ^2=1$, where $\alpha ^2=\alpha _{a}\alpha ^{a}$. In exponential form $h=e^{\beta ^{a}J_{a}}=\cosh \beta I+\frac{\beta ^{a}}{\beta}\sinh \beta J_{a}$ where $\beta ^2 =\beta _a\beta ^a$.

The field strength (or curvature) matrix valued two-form is $F=dA+A^2=F^{a}J_{a}={\bf F.\sigma }$.
It follows $F^{a}=dA^{a}-\epsilon ^{abc} A_{b}A_{c} $ or
$ {\bf F} = d{\bf A}- {\bf A\times A}$, where we define the "cross product" for differential forms of any order with indices in the adjoint representation of SO(2,1) as $(V\times W)^{a}=\epsilon ^{abc} V_{b}W_{c}$, while the "dot product" is $V^{a}W_{a}$.
If we have an algebra valued zero-form field $\alpha =\alpha ^{a}J_{a}={\bf \alpha .J}$ its covariant derivative is $D\alpha =D{\bf \alpha}.{\bf J}=d\alpha +[A,\alpha ]$ or $D{\bf \alpha}=d{\bf \alpha}+2{\bf \alpha\times A}$.

For the equation $A^h-A^{h^{-1}}=0$ we need $A^h=h^{-1}Ah+h^{-1}dh$. Considering $h^{-1}=\lambda I-\alpha ^{a}J_{a}$ and
using the properties of the generators, as in the SU(2) case, we get
$$h^{-1}Ah=[(\lambda ^2+\alpha ^2)A^{a}-2\lambda(A\times \alpha)^{a}-2({\bf A.\alpha)\alpha ^{a}}]J_{a}$$
and
$$h^{-1}dh=[\lambda d\alpha ^{a}-d\lambda \alpha ^{a}+(\alpha\times d\alpha)^{a}]J_{a}$$
We obtain then from the first equation
\begin{equation}
\lambda d\alpha ^{a}-d\lambda \alpha ^{a}=2\lambda (A\times \alpha)^{a}
\end{equation}
or
\begin{equation}
\lambda D\alpha ^{a}-d\lambda \alpha ^{a}=0
\end{equation}
The equation $F+F^h-\frac{1}{2}[A^h-A,A^h-A]=0$ yields
\begin{equation}
\lambda ^2F^{a}-\lambda (F\times\alpha)^{a}-({\bf F.\alpha})\alpha ^{a}+\frac{1}{2}[(\alpha\times D\alpha)\times(\alpha\times D\alpha)]^{a}=0
\end{equation}

\section{Solutions}

Here we will show that the gWZW model for SO(2,1) group, despite its simplicity, has non trivial black hole solutions. We found that the later have interesting properties, as we discuss below. We compute the Noether mass of the black hole and study its thermodynamics, finding that black holes of finite mass are massless and have zero entropy.

In solving the field equations
\begin{eqnarray}
D(h^2)=0\\
F+F^h-(h^{-1}Dh)^2=0
\end{eqnarray}
a remarkable simplification is achieved by setting $\lambda =0$ in $h=\lambda I+\alpha ^aJ_a$. In that case $h^2=-I$ and therefore the first of the previous field equations is fulfilled. Furthermore the second equation in the case $\lambda =0$ is
$$-(F\cdot\alpha)\alpha ^a+\frac{1}{2}[(\alpha\times D\alpha)\times(\alpha\times D\alpha)]^a=0$$
This is further simplified if we assume $F\cdot\alpha =0$, which is equivalent to $hF+Fh=0$, or $F=-h^{-1}Fh=-F^h$, which is a sort of anti-self-duality condition. We will use below an ansatz for which $F^0=F^1=0$, which follows from assuming that the torsion is zero, therefore the anti-self-duality condition $F\cdot\alpha =0$ would imply $\alpha ^2=0$.
Below we follow a ad hoc approach to the search of black hole solutions, but it turns out that the solutions of interest do indeed satisfy the conditions
\begin{eqnarray}
h^2=-I\\
F=-F^h
\end{eqnarray}

\subsection{Black holes}

Searching for generic black hole-like solutions leads to equations that we could not solve. Therefore we followed instead the strategy of assuming the metric to be the 1+1 black hole metric of Refs.\cite{CGHS-bh,witten-1+1-bh}, and look for configurations of the $\alpha$'s that would yield a solution of the full field equations of the theory.
The line element of the CGHS black hole \cite{CGHS-bh,witten-1+1-bh} is
\begin{equation}\label{metric}
ds^{2}=-\tanh(\gamma r)^{2}dt^{2}+dr^{2}
\end{equation}
where $\gamma$ is a constant. The vielbein is
\begin{eqnarray}\label{vielbein}
e^{0}=\tanh (\gamma r) dt~~~,~~~e^{1}=dr
\end{eqnarray}
and the torsion free spin connection $\omega^{01}$ is
\begin{equation}\label{omega}
\omega^{01}=\frac{\gamma}{\cosh^{2}(\gamma r)}dt.
\end{equation}
The Riemann curvature two form is then
\begin{equation}\label{R}
R^{01}=\frac{2\gamma^{2}\tanh(\gamma r)}{\cosh^{2}(\gamma r)}dt dr.
\end{equation}
and the curvature scalar is
$R=\frac{4\gamma^{2}}{\cosh^{2}(\gamma r)}$.

The idea now is to insert these ansatz into the field equations and see what equations must the $\alpha ^a$ satisfy.
As said before the $(A,F)$ are given in terms of the $(e,\omega)$, which in the torsion free case is
$A^{0}=\frac{1}{2}e^{1}~, ~A^{1}=\frac{1}{2}e^{0}~,~ A^{2}=-\frac{1}{2}\omega^{01}$
and $F^{0}=\frac{1}{2}T^{1}=0~,~ F^{1}=\frac{1}{2}T^{0}=0~,~F^{2}=-\frac{1}{2}(d\omega^{01}-e^{0}e^{1})$.

We could not solve the equations that result from that ansatz for generic $\alpha ^a$, but assuming time independence, as befits static solutions, and assuming that one of the components of $\alpha ^a$ vanish it is possible to find solutions (For detailed calculation of black hole solutions looking Appendix \ref{derivation_black_hole}).
There are no solutions with $\lambda \neq 0$.
If $\alpha^{2}=0$ and $\alpha^{0,1}(r,t)=\alpha^{0,1}(r)$ we get the solutions
\begin{equation}
\alpha^{0}=\pm\sqrt{(\alpha^{1})^{2}+1}~ ,~\alpha^{1}=\frac{C}{\tanh(\gamma r)}
\end{equation}
 where $C$ is a constant.\\

In the case $\alpha^{1}=0$ and $\alpha^{0,2}(r,t)=\alpha^{0,2}(r)$ we get the solution:\\
\begin{equation}
\alpha^{0}=\pm\sqrt{(\alpha^{2})^{2}+1}~,~\alpha^{1}=C \cosh(\gamma r)e^{\frac{1}{8\gamma^{2}}\cosh(2\gamma r)}
\end{equation}
where $C$ is a constant. For this second kind of black hole solutions we found, following methods similar to the ones applied in the next subsections to the first kind of black hole solutions, that their mass is infinite, which would arguably rule them out as sensible solutions.

\subsection{Black hole mass from Noether's theorem and a more general zero mass result}\label{noether_section}

Given a  generally covariant action $I[\phi ]$ for a physical theory, depending on a set of fields $\phi$, with  a  bulk lagrangian density $L$ and a boundary contribution, in a space-time $\Omega$ of the generic form

$$I=\int _{\Omega}L+\int _{\partial\Omega}B,$$

Noether's theorem states that the invariance under diffeomorphisms of this action implies that the following current is conserved
\begin{equation}
\star j=-\Theta - I_{\xi}L + d(I_{\xi}B)
\end{equation}
where the point dependent vector field $\xi$  generates the infinitesimal
diffeomorphisms $\delta x^{\mu}=\xi^{\mu}(x)$\footnote{The contraction operator $I_{\xi }$ is defined by
acting on a p-form $\alpha _p$ as
$$
I_{\xi }\alpha _{p}=\frac{1}{(p-1)!}\xi ^{\nu }\alpha _{\nu \mu
_{1}...\mu _{p-1}}dx^{\mu _{1}}...dx^{\mu _{p-1}}
$$
and being and anti-derivative in the sense that acting on the wedge
product of differential forms $\alpha _{p}$ and $\beta _{q}$ of
order p and q respectively gives $I_{\xi }(\alpha _{p}\beta
_{q})=I_{\xi }\alpha _{p}\beta _{q}+(-1)^{p}\alpha _{p}I_{\xi }\beta
_{q}$.}
and we can read $\Theta$ from
\[
\delta L=(Field~~~equations)\delta\phi+d\Theta
\]
where the variations $\delta\phi$ are infinitesimal but arbitrary in form. The Noether�s conserved charge $Q$ is defined as $Q=\int _{\mathcal{V}}\star j$, where $\mathcal{V}$ is a manifold that correspond to a fixed time slice of $\Omega$. The Noether's mass is the Noether's charge in the case $\xi=\frac{\partial ~}{\partial t}$.

We are interested in the Noether's mass of the black hole solution of the previous subsection. We have the 1+1 dimensional gWZW action (\ref{action})
$$
I_{gWZW}=\kappa\int_{\Sigma}{\frac{1}{3}\langle(h^{-1}dh)^{3}\rangle}-\kappa\int_{M^{2}}{\langle(A-h^{-1}dh)A^{h}\rangle},
$$
and we consider the $\alpha^{a}$�s and $A^{a}$�s corresponding to the black hole solution of eqs.(\ref{vielbein},\ref{omega}, \ref{R}).

We need to compute the different pieces of $\star j$. for our action there is no $B$ term, while from eq. (\ref{variation_action}) we can read $\Theta$ in arbitrary dimension
$$\Theta =-n(n+1)\{\int _0^1dt~t<\Delta A F_t^{n-1}h^{-1}\delta h>\}$$
which in our case ($n=1$) is
$$\Theta =-\frac{1}{2} <\Delta A~h^{-1}\delta h>$$
which vanishes because $\delta h =\mathcal{L}_{\xi}h=0$, where $\mathcal{L}_{\xi}h$ is the Lie derivative of $h$ along $\xi$, as $h$ is time independent.

We also have
$$I_{\xi}\langle(h^{-1}dh)^{3}\rangle =0$$
as there is no component of $\langle(h^{-1}dh)^{3}\rangle $ along $dt$ because $h$ is time independent.

Finally
\begin{eqnarray}\label{derivadasdelie}
I_{\xi}A&=&I_{\xi}A^{a}J_{a}=\frac{1}{2}\tanh(\gamma r)J_{1}-\frac{1}{2}\frac{\gamma}{\cosh^{2}(\gamma r)}J_{2} \nonumber\\
I_{\xi}A^{h}&=&I_{\xi}h^{-1}Ah=(-I_{\xi}A^{a}-2I_{\xi}A^{1}\alpha^{1}\alpha^{a}+2I_{\xi}A^{0}\alpha^{0}\alpha^{a})J_{a}\nonumber\\
            &=&-\tanh(\gamma r)\alpha^{1}\alpha^{0}J_{0}+(-\frac{1}{2}\tanh(\gamma r)-\tanh(\gamma r)\alpha^{1}\alpha^{1})J_{1}+\frac{1}{2}\frac{\gamma}{\cosh^{2}(\gamma r)}J_{2}\nonumber
\end{eqnarray}
We can now compute $I_{\xi}L$:
\begin{equation}
I_{\xi}L=I_{\xi}\langle(A-h^{-1}dh)A^{h}\rangle=\langle(I_{\xi}A)A^{h}\rangle-\langle(A-h^{-1}dh)I_{\xi}A^{h}\rangle,
\end{equation}
where we used that $h$ depends only on $r$ for this solution.  On the one hand
\begin{eqnarray}
\langle(I_{\xi}A)A^{h}\rangle=2(2\tanh(\gamma r)A^{0}\alpha^{0}\alpha^{1}-\frac{1}{2}(\alpha\otimes d\alpha)^{2}\frac{\gamma}{\cosh^{2}(\gamma r)}=\nonumber \\
=\tanh(\gamma r)\alpha^{0}\alpha^{1}dr-(\alpha\otimes d\alpha)^{2}\frac{\gamma}{\cosh^{2}(\gamma r)}.
\end{eqnarray}
and in the other hand
\begin{equation}
\langle(A-h^{-1}dh)I_{\xi}A^{h}\rangle=\tanh(\gamma r)\alpha^{0}\alpha^{1}dr-(\alpha\otimes d\alpha)^{2}\frac{\gamma}{\cosh^{2}(\gamma r)}.
\end{equation}
It follows that  $I_{\xi}\langle(A-h^{-1}dh)A^{h}\rangle$ vanishes, therefore, putting
all together, the Noether's mass of the black hole is zero.

It is in fact possible to prove that the Noether mass of a time independent configuration satisfying the ``anti-self-duality" condition $F=-F^h$ is zero. The $\Theta $ contribution vanishes as before because $h$ is time independent. Furthermore in computing $I_{\xi}L$ we can use the original expression for the transgression
$$I_{\xi}L=I_{\xi}2\int_0^1dt~<\Delta AF_t>=2~I_{\xi}<h^{-1}Dh[\frac{F+F^h}{2}-\frac{1}{6}(h^{-1}Dh)^2]>$$
If $F=-F^h$ the field equation reduces to $(h^{-1}Dh)^2=0$ and the previous expression vanishes,
implying that the Noether current is zero and so is the mass of that configuration.

\subsection{Black Hole thermodynamics}

\subsubsection{Temperature}

The CGHS line element is (see eq. (\ref{metric}))
$$ds^2=-\tanh ^2 (\gamma r)dt^2+dr^2$$
which for euclidean time $t_E=-it$ becomes the euclidean line element
$$ds_E^2=\tanh ^2 (\gamma r)dt_E^2+dr^2$$
The horizon of this black hole is at $r=0$.  The standard procedure to determine the Hawking temperature of a black hole is to study the near horizon geometry and chose the period $\beta$ of the euclidean time\footnote{The euclidean time period is related to the temperature $T$ by $\beta =\frac{1}{k_BT}$, where $k_B$ is the Boltzmann constant. In what follows we use natural units with $k_B=1$.} so that there is no conical singularity at the horizon. The reason for that is that the curvature diverges at the conical singularity what implies that that geometry is not a extremum of the euclidean action and will not be the one that contributes in a saddle point evaluation of the partition function.

Applying that procedure to the CGHS black hole geometry we have that the near horizon line element is
$$ds_E^2\cong (\gamma r)^2dt_E^2+dr^2=r^2 d\theta ^2+dr^2$$
where $\theta =\gamma t_E$. Avoiding a conical singularity would imply
$$\beta =\frac{2\pi }{\gamma}$$

The previous argument would apply for an action that explicitly includes the curvature $R^{ab}$, e.g. the Einstein-Hilbert action. However for the action (\ref{action}) only the spin connection appears, and not their derivatives, therefore we argue that a conical singularity at the horizon does not imply that the action is singular for that configuration and henceforth that the period $\beta$ of euclidean time is not restricted have a particular value. This situation is analogous to the case of extremal black holes.

\subsubsection{Euclidean Action}

In the semiclassical approximation of euclidean gravity the exponential of minus the euclidean action evaluated on configurations that extremize the action is proportional to the particion function
$$Z\approx N e^{-I_E}$$ and therefore the free energy $F$ is related to the euclidean action (\ref{action}) by $$\beta F= I_E+constant$$
We need therefore to evaluate the action
$$
I_{gWZW}=\kappa\int_{\Sigma}{\frac{1}{3}\langle(h^{-1}dh)^{3}\rangle}-\kappa\int_{M^{2}}{\langle(A-h^{-1}dh)A^{h}\rangle}
$$
on the black hole configuration.

The evaluation of the first integral of the second member would in principle require the extension of the two dimensional manifold $ M^2$ into the three dimensional manifold $\Sigma$. For instance we could picture the 1+1 black hole geometry as a "cigar-like" surface closing in $r=0$, with $r$ the coordinate along the cigar and $t_E$ the angular coordinate around the "cigar", and extend it to the solid interior of the "cigar" to carry on the 3D integration. However that would not be necessary, as the 1-form $h^{-1}dh$ has no component along $dt$, because $h$ is time independent, then the first term of the second member is zero
$$\int_{\Sigma}{\frac{1}{3}\langle(h^{-1}dh)^{3}\rangle}=0$$

In order to evaluate the second term, we have

\begin{eqnarray*}
{\langle(A-h^{-1}dh)A^{h}\rangle}&=&2\eta_{ab}(A^{a}-(\alpha\times d\alpha)^{a})(-A^{b}-2(\alpha\cdot A)+(\alpha\times d\alpha)^{b})\\
                                 &=&2\eta_{ab}(A^{a}-(\alpha\times d\alpha)^{a})(-2(\alpha\cdot A)\alpha^{b}-(A^{b}-(\alpha\times d\alpha)^{b}))\\
                                 &=&-4\eta_{ab}(A^{a}-(\alpha\times d\alpha)^{a})(\alpha\cdot A)\alpha^{b}\\
                                 &=&-4(\alpha\cdot A)(\alpha\cdot A)+4((\alpha\times d\alpha)\cdot\alpha)(\alpha\cdot A)
\end{eqnarray*}

The first term vanished due $\alpha\cdot A$ is a real valued 1-form while the second one vanished in virtue that $(\alpha\times d\alpha)\cdot\alpha=0$ if we replace the $\alpha^{a}$ of the black hole solution.
So the euclidean action is zero for the black hole solution.

Using the thermodynamic formulas $I_E=\beta F=S-\beta E$, where $S$ is the entropy and $E$ is the energy (mass), or equivalently
$M=E=-\frac{\partial I_E}{\partial \beta}$ and $S=I_E+\beta E$ we get $M=0$ and $S=0$.
It is reassuring that the thermodynamical mass agrees with the Noether�s mass computed in the previous section and is also vanishing.
The vanishing of the entropy, as the arbitrariness of the temperature, indicates that the black hole solutions considered here are extremal.

It can be shown that the euclidean action is zero for all configurations satisfying the ``anti-self-duality" condition $F=-F^h$. We have for the lagrangian density
$$L=2\int_0^1dt~<\Delta AF_t>=2<h^{-1}Dh[\frac{F+F^h}{2}-\frac{1}{6}(h^{-1}Dh)^2]>$$
If $F=-F^h$ the field equation is $(h^{-1}Dh)^2=0$ and the lagrangian density is zero, and so is the action.

\subsection{Discussion}

That the mass of the 1+1 black holes studied here should be zero struck us as odd at first, but further thought convinced us that in fact that result could even be considered natural, considering some heuristic arguments given below.

Massless gravitationally bound solutions have a rather long history in field theory. More than fifty years ago Pascual Jordan suggested that a star could be created from nothing as fas as its negative gravitational energy was exactly equal to its possitive rest mass\footnote{Einstein himself was so stunned when George Gamow told him about that idea that it almost did cost both of them a serious accident. As Gamow tells the story \cite{gamow}: "I remember that once, walking with him to the Institute, I mentioned Pascual Jordan's idea of how a star can be created from nothing, since at the point zero its negative gravitational mass defect is numerically equal to its positive rest mass. Einstein stopped in his tracks, and, since we were crossing a street several cars had to stop to avoid running us down.".}. This idea is in essence what is behind the creation {\it ex nihilo} of the universe scenarios discussed in modern cosmology. There are several examples of zero mass  black hole solutions in diverse dimensions and for different theories in the recent literature, for instance refs.\cite{lu,lu-pope}. It has been also claimed by Boulware et al.\cite{boulware} that for a specific gravitational theory, namely Conformal Gravity in a four dimensional space-time, every spatially bounded state must be massless.

As for the arguments that make the zero mass result plausible, firstly we could translate a heuristic argument made by Boulware et al.\cite{boulware} in their paper concerning the masslessness of bound states in conformal gravity (besides a formal argument) to the present context. The idea is that the action of conformal gravity in 4D has fourth order derivatives, what implies a propagator that goes as
$1/k^4$ in the four-momenta $k_{\mu}$, therefore the interaction or potential energy goes as $\int d^3k 1/k^4\sim 1/k\sim r$. Thus the potential of a localized source is linear and the work required to separate charge an infinite distance would be infinite, as happens with the interaction of quarks in QCD, which leads to instability against pair creation and color confinement on that case. Boulware et al. argue that conformal gravity must then be 'gravitationally confining', and as colour charge must be zero for bounded states in QCD, mass must be zero for bounded states in conformal gravity.

For a gravitational theory with second order derivatives in two space-time dimensions the propagator would go as $1/k^2$ and the potential would go as $\int dk
1/k^2\sim 1/k\sim r$, what would again imply that a generic gravitational theory with second order derivatives is gravitationally confining at the full quantum level and the mass of spatially localized states must be zero.
A similar argument was applied to the Schwinger model of electrodynamics with fermions in 1+1 dimensions \cite{schwinger} to argue that electrodynamics should be confining in that model, which was indeed confirmed by exact results.

These considerations seem to be in disagreement with the fact that CGHS black holes \cite{CGHS-bh,witten-1+1-bh} have non-zero masses.
However we observe that in the original article on CGHS black holes \cite{CGHS-bh} those black holes are proved to be unstable when quantum effects are taken in account, and in fact dubbed "evanescent black holes", and it is say there that the final state after their evaporation in the full quantum theory must be certain zero mass state. Similar considerations are made in ref.\cite{witten-1+1-bh}, where it is stated that the final state should be two copies of flat space.
The classical action considered in both refs.\cite{CGHS-bh,witten-1+1-bh} is different from the one considered here, however we may argue that because of the standard arguments for WZW actions \cite{witten-gWZW} the classical action considered here should receive no quantum corrections, being already the quantum effective action; and it could be even the quantum effective action for a whole class of classical gravitational actions in 1+1 dimensions. In that case a classical solution of the model studied here must have all quantum corrections included from the start, and therefore a static classical black hole solution must be, if the previous considerations hold, massless.

It is clear that the previous heuristic considerations must be taken with a grain of salt, as for instance the same argument applied to gravity in 3D would imply a logarithmic potential, that could also be confining, leading to only massless states, while the BTZ black holes of 3D gravity are certainly not massless in general, and one could also argue that the Chern-Simons gravity action in 3D does not receive quantum corrections. On the other hand one could argue that pure gravity in 3D has no propagating local degrees of freedom, therefore it makes no sense to speak of a propagator, making the Boulware et al. argument invalid in this case.

\section{Conclusions}

In this paper we derive the general field equations for topological gWZW models (gravitational or not) in any dimension and discuss the boundary contributions that could arise. In discussing gravitational gWZW models we consider the gauge group element $h$ to be generic, unlike refs.\cite{anabalon-1,anabalon-2} where it is chosen to have a specific form.

Another difference with refs.\cite{anabalon-1,anabalon-2} is that the gauge group considered in the 2D toy model we study is a de Sitter group in the dimension of the physical spacetime (2D) instead of a de Sitter group in the dimension that is reduced (that would be 3D for the toy model considered above), unlike what is done in the work of Anabalon et al.. That means that we only have one candidate vielbein, instead of two from which to arbitrarily chose, as they did. This choice is related to the fact that we had to chose as the required symmetrized trace the standard one, leading to Pontryagin-like invariants in 4D, instead of the Levi-Civita tensor, leading to Euler invariants in 4D, as Anabalon et al. did. As Pontryagin invariants exist in $d= 4 ~mod ~4$ dimensions,  then higher dimensional gWZW theories analogous to the 2D model considered here would exist in $d= 2 ~mod ~4$. That means that to obtain phenomenologically interesting 4D theories one must compactify from a 6D theory of this kind in the simplest case. In a way that is a drawback, as one would like to have a theory that is four dimensional by construction, but obtaining a 4D theory from a 6D theory by a sort of flux compactification could introduce a dynamically generated scale in the resulting theory, which would be desirable, as gWZW (as CS) theories have no dimensional constants in the action. At first glance a theory based on Pontryagin invariants seems to have the wrong parity, yet the presence of two intertwinned sets of fields makes it not so clear.

We regard the work presented here as a modest step towards the goal of investigating gravitational gWZW in higher dimensions, based either in the Pontryagin or Euler invariants, as a possible source for phenomenologically interesting models in 4D that go beyond General Relativity, and may in particular shed light on cosmological problems such as the origin of Dark Energy, or provide a gauge theoretic foundation for more ad hoc models, as for instance TeVeS.

In the framework of the 2D toy model we found black hole solutions with a metric of the CGHS type, which turned out to be massless (both from Noether's and thermodynamical analysis) and with zero entropy . It would be interesting to know how unique this solutions are, and to study their stability against small perturbations of the fields.\\

\begin{acknowledgements} We are grateful to Jorge Zanelli for conversations on the subject of this paper. P. Pais and P. Mora are thankful for hospitality at Centro de Estudios Cient\'{\i}ficos (CECs) at different times while working in this paper. Pablo Pais has been partially supported by a ANII Master's Degree Scholarship during this work. The Centro de Estudios Cient\'{\i}ficos (CECs) is funded by the Chilean Government through the Centers of Excellence Base Financing Program of Conicyt.
\end{acknowledgements}
\begin{appendix}

\section{Derivation of gWZW action field equations}\label{derivation_gwzw}

We compute the variation of the action and obtain the field equations in any dimensions starting from the formula for the variation of the transgression ,\cite{motz3}
$$
\delta\mathcal{T}_{2n+1}=(n+1) <F^n\delta A>-<\overline{F}^n\delta \overline{A}>
-n(n+1)d\{\int _0^1dt<\Delta AF_t^{n-1}\delta A_t>\}.
$$
If $\overline{A}=A^h=h^{-1}Ah+h^{-1}dh$ then $\overline{F}=h^{-1}Fh$ and $\delta \overline{A}=h^{-1}[\delta A +D(\delta h h^{-1})]h$, with the covariant derivative $D=d+[A,~~~]$. Then, using the Bianchi identity $DF=0$, the identity $d<something>=<D(something)>$, the cyclicity of the symmetrized trace and some algebra it results
\begin{equation}
\delta\mathcal{T}_{2n+1}=-(n+1)d\{ <F^n\delta h h^{-1}>
+n \int _0^1dt<\Delta AF_t^{n-1}\delta A_t>\}.
\end{equation}
Using $\Delta A =A-A^h$, $h\Delta A h^{-1}=A^{h^{-1}}-A\equiv -\tilde{\Delta A}$, $hF_th^{-1}=\tilde{F}_u$, where
the parameter $u$ is $u=1-t$ and $\tilde{F}_u=uF+(1-u)F^{h^{-1}}+u(u-1)\tilde{\Delta A}^2=d\tilde{A}_u+\tilde{A}_u^2$, with $\tilde{A}_u=uA+(1-u)A^{h^{-1}}$ and again with some algebra we obtain
\begin{eqnarray}
\delta\mathcal{T}_{2n+1}=-(n+1)d\{
n \int _0^1dt~t<[(A-A^h)F_t^{n-1}+(A^{h^{-1}}-A)\tilde{F}_t^{n-1}]\delta A>+\\ \nonumber
+\int _0^1dt<\tilde{F}_t^{n}\delta hh^{-1}>+n\int _0^1dt~t(t-1)<\tilde{\Delta A}\tilde{F}_t^{n-1}[\tilde{\Delta A},\delta hh^{-1}]>+\\ \nonumber
-d\{n\int _0^1dt~(1-t)<\tilde{\Delta A}\tilde{F}_t^{n-1}\delta hh^{-1}>\}\},
\end{eqnarray}
which can be rewritten as
\begin{eqnarray}
\delta\mathcal{T}_{2n+1}=-(n+1)d\{
n \int _0^1dt~t<[(A-A^h)F_t^{n-1}+(A^{h^{-1}}-A)\tilde{F}_t^{n-1}]\delta A>+\nonumber\\
+\int _0^1dt<F_t^{n}h^{-1}\delta h>+n\int _0^1dt~t(t-1)<\Delta A F_t^{n-1}[\Delta A,h^{-1}\delta h]>+\nonumber\\
-d\{n\int _0^1dt~(1-t)<\Delta A F_t^{n-1}h^{-1}\delta h>\}\}\nonumber.
\end{eqnarray}
The  variation o the gWZW action is therefore
\begin{eqnarray}
\delta I_{gWZW}=-(n+1)\int _{\Sigma ^{2n}}\{
n \int _0^1dt~t<[(A-A^h)F_t^{n-1}+(A^{h^{-1}}-A)\tilde{F}_t^{n-1}]\delta A>+\nonumber\\
+\int _0^1dt<F_t^{n}h^{-1}\delta h>+n\int _0^1dt~t(t-1)<\Delta A F_t^{n-1}[\Delta A,h^{-1}\delta h]>\}+\\ \nonumber
+n(n+1)\int _{\partial\Sigma ^{2n}}\{\int _0^1dt~(1-t)<\Delta A F_t^{n-1}h^{-1}\delta h>\}\nonumber,
\end{eqnarray}
where the space-time manifold is $\Sigma ^{2n}\equiv\partial\mathcal{M}$ and $\partial\Sigma ^{2n}$ is its $2n-1$-dimensional boundary.

The field equations derived from the action principle $\delta I_{gWZW}=0$, with the ones corresponding to $\delta A$ being
\begin{eqnarray}
\int _0^1dt~t<[(A-A^h)F_t^{n-1}+(A^{h^{-1}}-A)\tilde{F}_t^{n-1}]G^A>=0,
\end{eqnarray}
and the ones corresponding to $\delta h$ being
\begin{eqnarray}
 \int _0^1dt<F_t^{n}G^A>+n\int _0^1dt~t(t-1)<\Delta A F_t^{n-1}[\Delta A,G^A]>=0
\end{eqnarray}

\section{Derivation of black hole solutions}\label{derivation_black_hole}

The line element of the CGHS black hole \cite{CGHS-bh,witten-1+1-bh} is (\ref{metric})
\begin{equation}
ds^{2}=-\tanh(\gamma r)^{2}dt^{2}+dr^{2},\nonumber
\end{equation}
where $\gamma$ is a constant.

The idea now is to insert these ansatz into the field equations and see what equations must the $\alpha ^a$ satisfy. As said before the $(A,F)$ are given in terms of the $(e,\omega)$, which in the torsion free case is
$A^{0}=\frac{1}{2}e^{1}~, ~A^{1}=\frac{1}{2}e^{0}~,~ A^{2}=-\frac{1}{2}\omega^{01}$
and $F^{0}=\frac{1}{2}T^{1}=0~,~ F^{1}=\frac{1}{2}T^{0}=0~,~F^{2}=-\frac{1}{2}(d\omega^{01}-e^{0}e^{1})$, with the vielbein and spin connection of Section 4.3.1.

\subsection{Case $\lambda \neq 0$}

We can write the field equations for each value of $a=0,1,2$ and separate components along $dr$ and $dt$.

For the first field equation and for each component we have:
\begin{eqnarray}
(\alpha_{a}\alpha^{a}+1)\left(\dot{\alpha^{0}}+\frac{\gamma\alpha^{1}}{\cosh^{2}(\gamma r)}+\tanh (\gamma r)\alpha^{2}\right) - \alpha^{0}(\alpha^{1}\dot{\alpha^{1}}+\alpha^{2}\dot{\alpha^{2}}-\alpha^{0}\dot{\alpha^{0}})=0\nonumber\\
(\alpha_{a}\alpha^{a}+1)\alpha^{0'}-\alpha^{0}(\alpha^{1}\alpha^{1'}+\alpha^{2}\alpha^{2'}-\alpha^{0}\alpha^{0'})=0\nonumber\\
(\alpha_{\mu}\alpha^{\mu}+1)\left(\dot{\alpha^{1}}+\frac{\gamma\alpha^{0}}{\cosh^{2}r}\right) - \alpha^{1}(\alpha^{1}\dot{\alpha^{1}}+\alpha^{2}\dot{\alpha^{2}}-\alpha^{0}\dot{\alpha^{0}})=0\nonumber\\
(\alpha_{a}\alpha^{a}+1)(\alpha^{1'}+\alpha^{2})-\alpha^{1}(\alpha^{1}\alpha^{1'}+\alpha^{2}\alpha^{2'}-\alpha^{0}\alpha^{0'})=0\nonumber\\
(\alpha_{a}\alpha^{a}+1)\left(\dot{\alpha^{2}}+\tanh(\gamma r)\alpha^{0}\right) - \alpha^{2}(\alpha^{1}\dot{\alpha^{1}}+\alpha^{2}\dot{\alpha^{2}}-\alpha^{0}\dot{\alpha^{0}})=0\nonumber\\
(\alpha_{a}\alpha^{a}+1)(\alpha^{2'}-\alpha^{1})-\alpha^{1}(\alpha^{1}\alpha^{1'}+\alpha^{2}\alpha^{2'}-\alpha^{0}\alpha^{0'})=0\nonumber,
\end{eqnarray}
where $\dot{\alpha^{a}}=\frac{\partial\alpha^{a}}{\partial t}$ y $\alpha^{a'}=\frac{\partial\alpha^{\mu}}{\partial r}$.
Using that if $\lambda\neq 0$ the first field equation implies $D\alpha^{a}=\frac{d\lambda}{\lambda}\alpha^{a}$, we get
\begin{equation}
(\alpha\times D\alpha)^{a}=\epsilon_{bcd}\eta^{da}\alpha^{b}D\alpha^{c}=\epsilon_{bcd}\eta^{da}\alpha^{b}\frac{d\lambda}{\lambda}\alpha^{c}=0.
\end{equation}
The second field equation reduces then to
\begin{equation}
\lambda^{2}F^{a} - \lambda(F\times\alpha)^{a} - (F.\alpha)\alpha^{a} = 0.
\end{equation}
Using our ansatz for $A$ we get, for each component:
\begin{eqnarray}
\alpha^{2}\alpha^{0}+\sqrt{\alpha_{a}\alpha^{a}+1}\alpha^{1}=0\\
\alpha^{2}\alpha^{1}+\sqrt{\alpha_{a}\alpha^{a}+1}\alpha^{0}=0\\
\lambda^{2}-\alpha^{2}\alpha^{2}=0.
\end{eqnarray}
Using Maple12 the following complex solutions result:\\
A. $\alpha^{0}=\alpha^{0}~,~\alpha^{1}=-i~,~\alpha^{2}=0$\\
B. $\alpha^{0}=\alpha^{0}~,~\alpha^{1}=i~,~\alpha^{2}=0$,\\
which are not allowed because $\alpha^{a}$ must be real, and furthermore $\lambda^{2}=(\alpha^{1})^{2}+1=0$, against our assumption.

Other solutions are:\\
C. $\alpha^{0}=1~,~\alpha^{1}=0~,~\alpha^{2}=0$\\
D. $\alpha^{0}=-1~,~\alpha^{1}=0~,~\alpha^{2}=0$
which against have  $\lambda^{2}=-(\alpha^{0})^{2}+1=0$.

Finally a solution is:\\
E. $(\alpha^{1})^{2}=(\alpha^{2})^{2}-1~,~\dot{\alpha^{1}}=-\frac{\alpha^{1}\alpha^{0}\dot{\alpha^{0}}}{1-(\alpha^{0})^{2}}~,~\alpha^{1'}=\frac{\alpha^{1}\alpha^{0}\alpha^{0'}}{-1+(\alpha^{0})^{2}}~,~
\alpha^{2}=0$
which also has $\lambda^{2}=-(\alpha^{0})^{2}+(\alpha^{1})^{2}+1=0$.\\
In conclusion the solutions obtained using Maple12 are with $\lambda=0$,
againt the initial assumption of $\lambda\neq 0$, and must be discarded.

\subsection{Case $\lambda=0$}

In this case Field Equation I
$\lambda d\alpha^{a} - d\lambda\alpha^{a} - 2\lambda(A\times\alpha)^{a} =0$
is trivially  satisfied if $\lambda=0$.

Field Equation II is

\begin{equation}
- 2(F.\alpha)\alpha^{a} -  ((\alpha\times D\alpha)\times(\alpha\times D\alpha))^{a} = 0.
\end{equation}

Using $(\alpha\times D\alpha)^{a}=\epsilon_{bcd}\eta^{da}\alpha^{b}D\alpha^{c}$ and
$((\alpha\times D\alpha)\times((\alpha\times D\alpha))^{a}=\epsilon_{bcd}\eta^{da}(\alpha\times D\alpha)^{b}(\alpha\times D\alpha)^{c}$
we obtain

\begin{eqnarray}
(\alpha^{0}D\alpha^{1}-\alpha^{1}D\alpha^{0})(\alpha^{2}D\alpha^{0}-\alpha^{0}D\alpha^{2})-F^{2}\alpha^{2}\alpha^{0}=0\\
(\alpha^{0}D\alpha^{1}-\alpha^{1}D\alpha^{0})(\alpha^{2}D\alpha^{1}-\alpha^{1}D\alpha^{2})-F^{2}\alpha^{2}\alpha^{1}=0\\
(\alpha^{2}D\alpha^{1}-\alpha^{1}D\alpha^{2})(\alpha^{2}D\alpha^{0}-\alpha^{0}D\alpha^{2})-F^{2}\alpha^{2}\alpha^{2}=0.
\end{eqnarray}
From the field equations and the ansatz for $A$ we obtain the system of equations
\begin{eqnarray}
\frac{\alpha^{2}\alpha^{0}\tanh(r)}{2}\left(-\frac{2\gamma^{2}}{\cosh(\gamma r)^{2}}+1\right)=\nonumber\\
=\left(\alpha^{0}\dot{\alpha^{1}}+\frac{\gamma\alpha^{0}\alpha^{0}}{\cosh(\gamma r)^{2}}-\alpha^{1}\dot{\alpha^{0}}-\alpha^{1}\alpha^{2}\tanh(\gamma r)-\frac{\gamma\alpha^{1}\alpha^{1}}{\cosh(\gamma r)^{2}}\right)\times\nonumber\\ \times (\alpha^{2}\alpha^{0'}-\alpha^{0}\alpha^{2'}+\alpha^{0}\alpha^{1})-\nonumber\\
-(\alpha^{0}\alpha^{1'}+\alpha^{0}\alpha^{2}-\alpha^{1}\alpha^{0'})\times\nonumber\\ \times\left(\alpha^{2}\dot{\alpha^{0}}+\alpha^{2}\alpha^{2}\tanh(\gamma r)+\frac{\gamma\alpha^{2}\alpha^{1}}{\cosh(\gamma r)^{2}}-\alpha^{0}\dot{\alpha^{2}}-\alpha^{0}\alpha^{0}\tanh(\gamma r)\right)\nonumber
\end{eqnarray}

\begin{eqnarray}
\frac{\alpha^{2}\alpha^{1}\tanh(\gamma r)}{2}\left(-\frac{2\gamma^{2}}{\cosh(\gamma r)^{2}}+1\right)=\nonumber\\
=\left(\alpha^{0}\dot{\alpha^{1}}+\frac{\gamma\alpha^{0}\alpha^{0}}{\cosh(\gamma r)^{2}}-\alpha^{1}\dot{\alpha^{0}}-\alpha^{1}\alpha^{2}\tanh(\gamma r)-\frac{\gamma\alpha^{1}\alpha^{1}}{\cosh(\gamma r)^{2}}\right)\times\nonumber\\
\times(\alpha^{2}\alpha^{1'}+\alpha^{2}\alpha^{2}-\alpha^{1}\alpha^{2'}+\alpha^{1}\alpha^{1})-\nonumber\\-(\alpha^{0}\alpha{1'}+\alpha^{0}\alpha^{2}-\alpha^{1}\alpha^{0'})\left(\alpha^{2}\dot{\alpha^{1}}+\frac{\gamma\alpha^{2}\alpha^{0}}{\cosh(\gamma r)^{2}}-\alpha^{1}\dot{\alpha^{2}}-\alpha^{1}\alpha^{0}\tanh(\gamma r)\right)\nonumber
\end{eqnarray}

\begin{eqnarray}
\frac{\alpha^{2}\alpha^{2}\tanh(\gamma r)}{2}\left(-\frac{2\gamma^{2}}{\cosh(\gamma r)^{2}}+1\right)=\nonumber\\
=\left(\alpha^{2}\dot{\alpha^{1}}+\frac{\gamma\alpha^{2}\alpha^{0}}{\cosh(\gamma r)^{2}}-\alpha^{1}\dot{\alpha^{2}}-\alpha^{1}\alpha^{0}\tanh(\gamma r)\right)\times\nonumber\\ \times(\alpha^{2}\alpha^{0'}-\alpha^{0}\alpha^{2'}+\alpha^{0}\alpha^{1})-\nonumber\\
-(\alpha^{2}\alpha^{1'}+\alpha^{2}\alpha^{2}-\alpha^{1}\alpha^{2'}+\alpha^{1}\alpha^{1})\times\nonumber\\ \times\left(\alpha^{2}\dot{\alpha^{0}}+\alpha^{2}\alpha^{2}\tanh(\gamma r)+\frac{\gamma\alpha^{2}\alpha^{1}}{\cosh(\gamma r)^{2}}-\alpha^{0}\dot{\alpha^{2}}-\alpha^{0}\alpha^{0}\tanh(\gamma r)\right)\nonumber\\
(\alpha^{1})^{2}+(\alpha^{2})^{2}-(\alpha^{0})^{2}+1=0 \nonumber,
\end{eqnarray}

We could not solve this equations for generic $\alpha ^a$, but assuming time independence, as befits a static solution, and assuming one of the components of $\alpha ^a$ vanish it is possible to find the following solutions:

If $\alpha^{2}=0$ and $\alpha^{0,1}(r,t)=\alpha^{0,1}(r)$ we get
$$\alpha^{0}=\pm\sqrt{(\alpha^{1})^{2}+1}~ ,~\alpha^{1}=\frac{C}{\tanh(\gamma r)}$$
where $C$ is a constant.\\

If $\alpha^{1}=0$ and $\alpha^{0,2}(r,t)=\alpha^{0,2}(r)$ we get
$$\alpha^{0}=\pm\sqrt{(\alpha^{2})^{2}+1}~,~\alpha^{1}=C \cosh(\gamma r)e^{\frac{1}{8\gamma^{2}}\cosh(2\gamma r)}$$
where $C$ is a constant.
\end{appendix}

\end{document}